\documentclass[twocolumn,twocolappendix,times,tighten]{aastex631}
\usepackage{multirow}

\newcommand{\Mj}{$M_{\rm Jup}$}  

\shorttitle{Monitoring Accretion onto FU Tau B}
\shortauthors{Wu et al.}
  
\published{in AJ}
  
\begin{document}
\title{{\large Monitoring H$\alpha$ Emission from the Wide-orbit Brown-dwarf Companion FU Tau B}}

\author[0000-0002-4392-1446]{Ya-Lin Wu}
\affiliation{\textup{Department of Physics, National Taiwan Normal University, Taipei 116, Taiwan; {\textcolor{teal}{yalinwu@ntnu.edu.tw}}}}
\affiliation{\textup{Center of Astronomy and Gravitation, National Taiwan Normal University, Taipei 116, Taiwan}}

\author{Yu-Chi Cheng}
\affiliation{\textup{Department of Physics, National Taiwan Normal University, Taipei 116, Taiwan; {\textcolor{teal}{yalinwu@ntnu.edu.tw}}}}
\affiliation{\textup{Center of Astronomy and Gravitation, National Taiwan Normal University, Taipei 116, Taiwan}}

\author[0000-0002-9679-5279]{Li-Ching Huang}
\affiliation{\textup{Department of Physics, National Taiwan Normal University, Taipei 116, Taiwan; {\textcolor{teal}{yalinwu@ntnu.edu.tw}}}}
\affiliation{\textup{Center of Astronomy and Gravitation, National Taiwan Normal University, Taipei 116, Taiwan}}

\author[0000-0003-2649-2288]{Brendan P. Bowler}
\affiliation{\textup{Department of Astronomy, The University of Texas at Austin, Austin, TX 78712, USA}}

\author[0000-0002-2167-8246]{Laird M. Close}
\affiliation{\textup{Steward Observatory, University of Arizona, Tucson, AZ 85721, USA}}

\author[0000-0001-7322-3801]{Wei-Ling Tseng}
\affiliation{\textup{Center of Astronomy and Gravitation, National Taiwan Normal University, Taipei 116, Taiwan}}
\affiliation{\textup{Department of Earth Sciences, National Taiwan Normal University, Taipei 116, Taiwan}}

\author{Ning Chen}
\affiliation{\textup{Department of Physics, National Taiwan Normal University, Taipei 116, Taiwan; {\textcolor{teal}{yalinwu@ntnu.edu.tw}}}}

\author{Da-Wei Chen}
\affiliation{\textup{Graduate Institute of Physics, National Taiwan University, Taipei 106, Taiwan}}

\begin{abstract} 
\noindent Monitoring mass accretion onto substellar objects provides insights into the geometry of the accretion flows. We use the Lulin One-meter Telescope to monitor H$\alpha$ emission from FU Tau B, a $\sim$19 \Mj\,brown-dwarf companion at 5\farcs7 (719 au) from the host star, for six consecutive nights. This is the longest continuous H$\alpha$ monitoring for a substellar companion near the deuterium-burning limit. We aim to investigate if accretion near the planetary regime could be rotationally modulated as suggested by magnetospheric accretion models. We find tentative evidence that H$\alpha$ mildly varies on hourly and daily timescales, though our sensitivity is not sufficient to definitively establish any rotational modulation. No burst-like events are detected, implying that accretion onto FU Tau B is overall stable during the time baseline and sampling windows over which it was observed. The primary star FU Tau A also exhibits H$\alpha$ variations over timescales from minutes to days. This program highlights the potential of monitoring accretion onto substellar objects with small telescopes.\vspace{7pt} \\
{\it Unified Astronomy Thesaurus concepts:} Accretion (14); Brown dwarfs (185); Stellar accretion (1578); Time series analysis (1916); Lomb--Scargle periodogram (1959)\\  
\end{abstract}

\section{Introduction}
\label{sect1}
Variability of mass accretion onto giant planets provides clues to their formation timescales and the geometry of the accretion flow. In magnetospheric accretion models (e.g., \citealt{Konigl91,Batygin18,Thanathibodee19}), mass inflow along the nearly pole-on magnetic field lines could form hot spots that corotate with the planet, making the shock-induced emission lines potentially variable over the rotation period. On the other hand, unsteady inflow and obscuration from the tilted or puffed inner disk could add stochastic bursts and dips to the rotational modulation (e.g., \citealt{Bouvier99,Cody14}). It is thus desirable to measure accretion variability on timescales relevant to these phenomena and mechanisms.

While there has been accretion monitoring for T Tauri stars (e.g., \citealt{Nguyen09,Biazzo12,Pouilly20,Sousa21,Zsidi22}) and young isolated brown dwarfs (e.g., \citealt{Natta04,SJ06,Stelzer07,Herczeg09}), such efforts near the planetary regime have been rare. Among the dozens of young substellar companions and protoplanets discovered in direct-imaging surveys, very few of them have multiepoch accretion-rate measurements (e.g., GQ Lup B and GSC 06214--00210 B, \citealt{Demars23}; PDS 70 b, \citealt{Zhou21}). Moreover, many of these studies were carried out at different observatories, further complicating the interpretations. A monitoring program with identical instrumentation and data reduction is beneficial to measure the mean accretion rate and variability and reduce systematics. This is important to constrain the slope and scatter of the relationship between object mass and accretion rate in the substellar regime, which may reflect the formation mechanism for these wide companions. For instance, \cite{SH15} suggested that companions formed via disk fragmentation should tend to be more actively accreting than those formed in collapsing prestellar cores. Constraining mass accretion also helps estimate the dissipation timescale of circumsubstellar disks and therefore the growth timescale of these wide companions and their satellites (e.g., \citealt{Benisty21,Wu22}).

At optical wavelengths, the H$\alpha$ emission at 6563\,\AA \,is arguably the best accretion tracer as it is the most prominent hydrogen recombination line---often $\mathcal{O}(10^2)$ brighter than the photosphere and the shock-induced continuum excess. Indeed, several young brown-dwarf companions and protoplanets have strong H$\alpha$ emission indicative of active accretion (e.g., \citealt{Zhou14,Zhou21,SantamariaMiranda18,Wagner18,Eriksson20}). Follow-up H$\alpha$ monitoring opens the possibility of probing the variability amplitude and periodicity. As wide substellar companions are typically hundreds to thousands of au from their hosts, it is possible to resolve the widest pairs under moderate seeing without resorting to adaptive optics systems. Small telescopes can play an important and complementary role in monitoring accretion in these systems alongside large ground-based and space facilities.

\begin{figure}[t]
\centering
\figurenum{1}
\includegraphics[width=\linewidth]{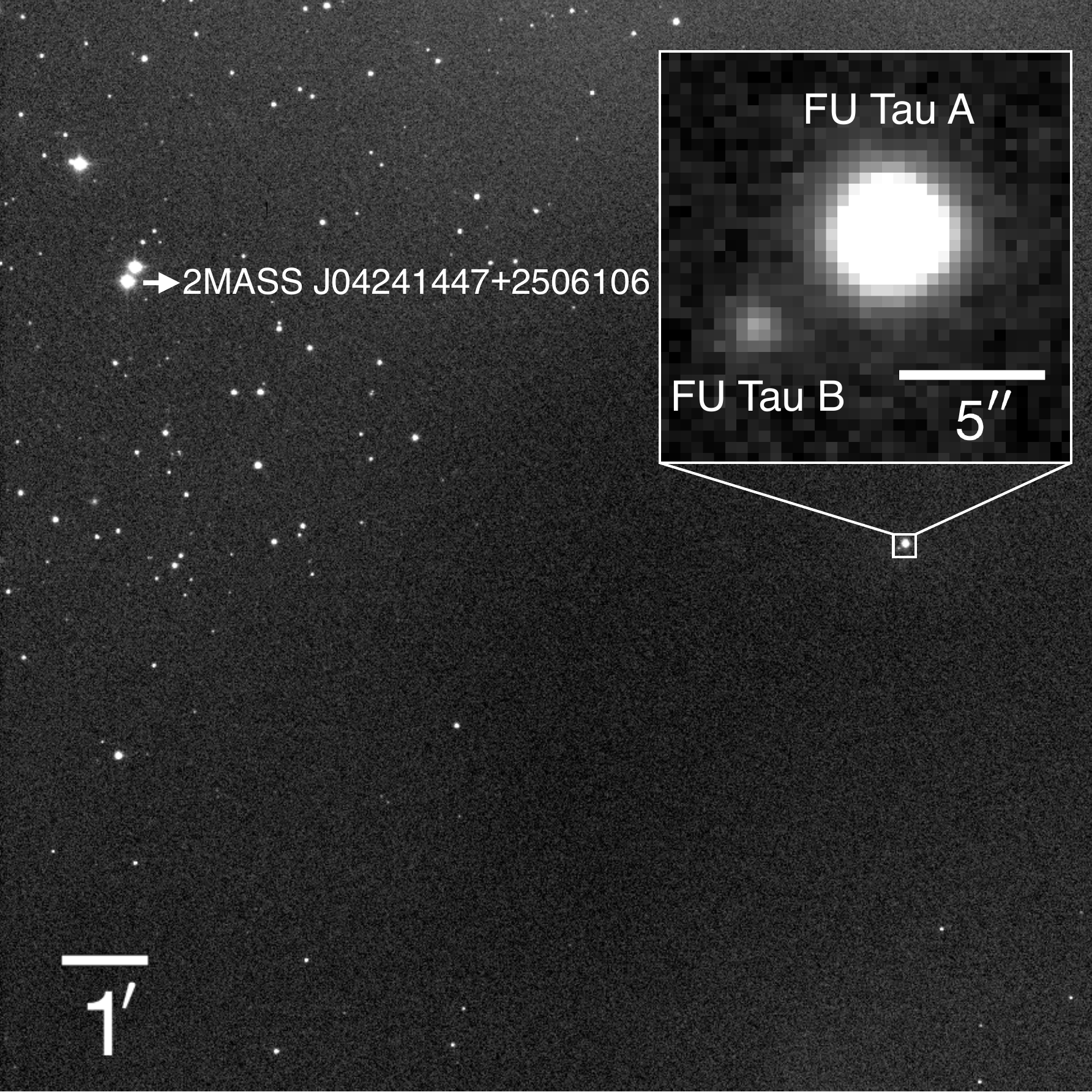}\vspace{-3pt}
\caption{The FU Tau system and the neighboring field stars. This image was made by stacking all frames in December 6. The Barnard 215 dark cloud \citep{Barnard27} obscures many field stars from the central to the lower-right regions. 2MASS J04241447$+$2506106 is our H$\alpha$ reference star. North is up and east is left.}
\label{fig:FOV}\vspace{-7pt}
\end{figure}

Here we present our six-night H$\alpha$ imaging of the FU Tau system with the Lulin One-meter Telescope (LOT) at Lulin Observatory in Taiwan. This is the longest continuous monitoring of an accretion-tracing emission line for a brown-dwarf companion. The primary star\footnote{The mass of FU Tau A can be sensitive to the adopted physical parameters. In Section \ref{sub_A}, we show that it might be a low-mass star based on the revised parameters in \cite{Bowler23}.} FU Tau A is actively accreting \citep{Luhman09,Stelzer10,Stelzer13,Rodriguez17}, and its disk was imaged with the Atacama Large Millimeter/submillimeter Array (ALMA) in 0.88 mm dust continuum \citep{Wu20}. \cite{Bowler23} recently determine a stellar inclination of $i_\ast=75^{+14}_{-5}\degr$ based on $v$ sin $i_\ast=17.4\pm0.3$ km s$^{-1}$ and the rotation period of 3.93 d derived from the observations of the Transiting Exoplanet Survey Satellite (TESS). The wide-orbit brown-dwarf companion FU Tau B, discovered by \cite{Luhman09}, is at 5$\farcs$69 and a position angle of $122\fdg8$ \citep{Todorov14} from the primary (projected separation of 719 au at a Gaia DR3 distance of $126.4\pm1.0$ pc; \citealt{Bailer-Jones21}). \cite{Luhman09} also detected its strong H$\alpha$ emission (equivalent width $\sim$70\,\AA) and near-infrared excess, implying that FU Tau B also harbors an accretion disk. In Section \ref{sub_B}, we estimate a mass of $\sim$19 \Mj \,for FU Tau B using the evolutionary models in \cite{Baraffe15}.
\vspace{-10pt}

\begin{deluxetable}{cccc}
\tablewidth{\columnwidth}
\tablecaption{Lulin H$\alpha$ Monitoring of FU Tau \label{tb1}}
\tablehead{
\colhead{\hspace{.2cm}UT Date}\hspace{.2cm} &
\colhead{\hspace{.2cm}Exposure}\hspace{.2cm} &
\colhead{\hspace{.2cm}$N_{\rm frame}$\tablenotemark{{\footnotesize a}}}\hspace{.2cm} &
\colhead{\hspace{.2cm}Average Seeing}\hspace{.2cm}  
}
\startdata 
2021-12-04 	& 600 s	&	27 (22)		& 	$1\farcs4$  \\
2021-12-05 	& 300 s	&	81 (72)		& 	$1\farcs9$  \\
2021-12-06 	& 300 s	&	83 (73)		& 	$1\farcs3$  \\
2021-12-07 	& 300 s	&	69 (60)		& 	$2\farcs1$  \\ 
2021-12-08 	& 300 s	&	100 (90)		& 	$1\farcs2$  \\
2021-12-09	& 300 s	&	111 (96)		& 	$1\farcs2$
\enddata
\tablenotetext{}{{\bf Note.} {\normalsize $^a$}\,Total number of frames and those taken at elevations of $>$30\degr\,(in parentheses).} \vspace{-21pt}
\end{deluxetable}

\section{Observations and Data Reduction}
We monitored the FU Tau system at H$\alpha$ with the LOT on UT 2021 December 4--9. The central wavelength and the effective width of the H$\alpha$ filter are 6562.8\,\AA \,and 36.7\,\AA, respectively. The Lulin CCD camera has a pixel scale of $0\farcs383$ and a field of view of $13\farcm07$\,$\times$\,$13\farcm07$. Figure \ref{fig:FOV} shows the typical field of our observations. The weather on these nights was photometric, with average seeing ranging from $1\farcs2$ to $2\farcs1$. We kept imaging FU Tau as long as it was above the telescope elevation limit of 20$\degr$. The integration time for individual frames is 600\,s for the first night and 300\,s for the subsequent nights. Table \ref{tb1} summarizes our observations. 

\begin{figure*}[t]
\centering
\figurenum{2}
\includegraphics[width=\linewidth]{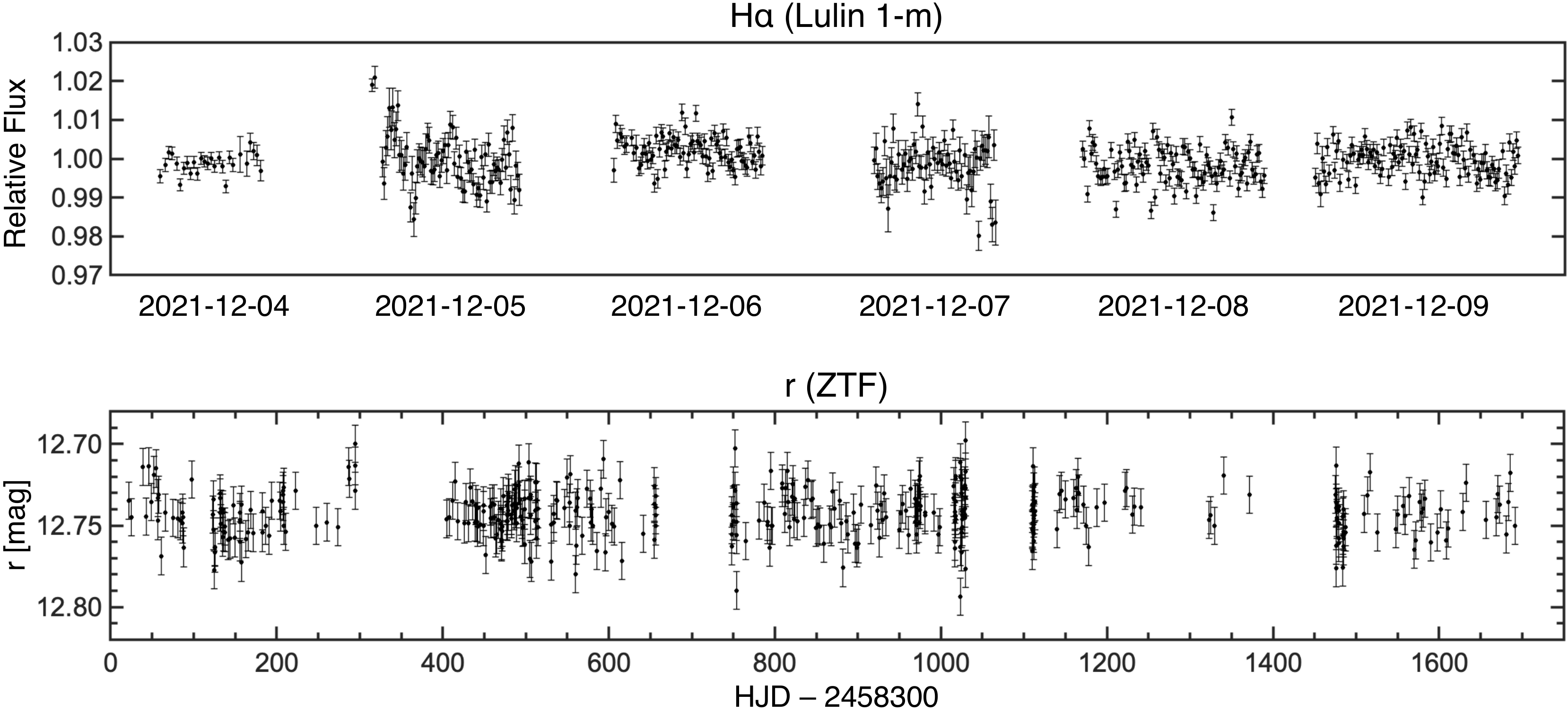} \vspace{-17pt}
\caption{Top: the calibrated light curve of 2MASS J04241447$+$2506106 indicates that it only has 0.3\%--0.7\% variations at H$\alpha$ over the course of our observations. Bottom: the ZTF $r$-band monitoring over 1670 days shows that 2MASS J04241447$+$2506106 is stable to a 1.4\% level.}\vspace{2pt}
\label{fig:refs}
\end{figure*}

The raw data were dark subtracted, flat corrected, and cross correlated with respect to FU Tau A. As we did not observe a known optical standard star, we searched for a suitable H$\alpha$ photometric reference from more than 30 field stars\footnote{The number of field stars involved in this iterative process is different each night (36, 32, 37, 37, 34, and 33) due to the slightly different telescope pointings.} following the iterative approach in \cite{Radigan12}. In brief, for each reference star candidate, a corrected light curve was computed by dividing its raw light curve by a calibration curve, which was created by median-combining the normalized light curves of the other candidates. Stars with high standard deviations in their corrected light curves were discarded. We iterated the above process and found that 2MASS J04241447$+$2506106 (see Figure \ref{fig:FOV}) was the most stable at H$\alpha$ over six nights, with standard deviations of 0.3\%, 0.7\%, 0.4\%, 0.6\%, 0.5\%, and 0.4\%, respectively. The top panel of Figure \ref{fig:refs} displays its calibrated H$\alpha$ light curve. To further understand if it could be chromospherically active, we examined the 1670 day $r$-band data from the Zwicky Transient Facility (ZTF; \citealt{Bellm19,Masci19}), as shown in the bottom panel of Figure \ref{fig:refs}. This long-term monitoring demonstrates that 2MASS J04241447$+$2506106 is likely stable. We thus measured $\Delta$H$\alpha$, the magnitude difference at H$\alpha$, between FU Tau A and 2MASS J04241447$+$2506106 in each frame and listed the results in Table \ref{tb2}.  \vspace{-22pt}

\begin{deluxetable}{ccc}
\tablecaption{Relative Photometry between FU Tau A and 2MASS J04241447$+$2506106\label{tb2}}
\tablehead{
\colhead{\hspace{.82cm} Date}\hspace{.82cm}  &
\colhead{\hspace{.82cm} $\Delta{\rm H}_\alpha$}\hspace{.82cm}  &
\colhead{\hspace{.82cm} $\sigma_{\Delta{\rm H}\alpha}$}\hspace{.82cm}   
}
\startdata 
2459553.15508 & 1.900 & 0.005 \\ 
2459553.16581 & 1.897 & 0.005 \\ 
2459553.17281 & 1.897 & 0.005 \\ 
$\cdots$ & $\cdots$ & $\cdots$ \\
2459558.34847 & 1.945 & 0.008 
\enddata
\tablenotetext{}{{\bf Notes.} (1) Heliocentric Julian date at the exposure midpoint. (2) Magnitude difference at H$\alpha$. (3) Uncertainty of the magnitude difference at H$\alpha$. (This table is available in its entirety in machine-readable form.)} \vspace{-32pt}
\end{deluxetable}

\begin{deluxetable}{ccc}
\tablecaption{Relative Photometry between FU Tau B and 2MASS J04241447$+$2506106\label{tb3}}
\tablehead{
\colhead{\hspace{.82cm} Date}\hspace{.82cm}  &
\colhead{\hspace{.82cm} $\Delta{\rm H}_\alpha$}\hspace{.82cm}  &
\colhead{\hspace{.82cm} $\sigma_{\Delta{\rm H}\alpha}$}\hspace{.82cm}  
}
\startdata 
2459553.16456 & 7.07 & 0.12 \\ 
2459553.18894 & 6.92 & 0.08 \\ 
2459553.21100 & 6.97 & 0.09 \\ 
$\cdots$ & $\cdots$ & $\cdots$ \\
2459558.30415 & 6.98 & 0.14
\enddata
\tablenotetext{}{{\bf Notes.} (1) Mean heliocentric Julian date. (2) Magnitude difference at H$\alpha$. (3) Uncertainty of the magnitude difference at H$\alpha$. (This table is available in its entirety in machine-readable form.)} \vspace{-32pt}
\end{deluxetable}

To accurately measure the flux density of the companion, we need to remove the primary star. As there were no suitable stars to serve as a point spread function (PSF) template in our data, we subtracted the best-fit elliptical Moffat function derived with the Levenberg--Marquardt algorithm at the position of FU Tau A. To better bring out the companion and enhance the signal-to-noise ratios (S/Ns), we limited our analysis to data with elevations of $>$30$\degr$ (the number of frames in parentheses in Table \ref{tb1}). We median combined these PSF-subtracted frames into 30 minutes bins for those taken on UT December 4, 5, 6, 8, 9, and 1 hr bins for data taken on UT December 7 due to poor seeing. Relative photometry between FU Tau B and 2MASS J04241447$+$2506106 is listed in Table \ref{tb3}.

\begin{figure*}[h]
\centering
\figurenum{3}
\includegraphics[width=\linewidth]{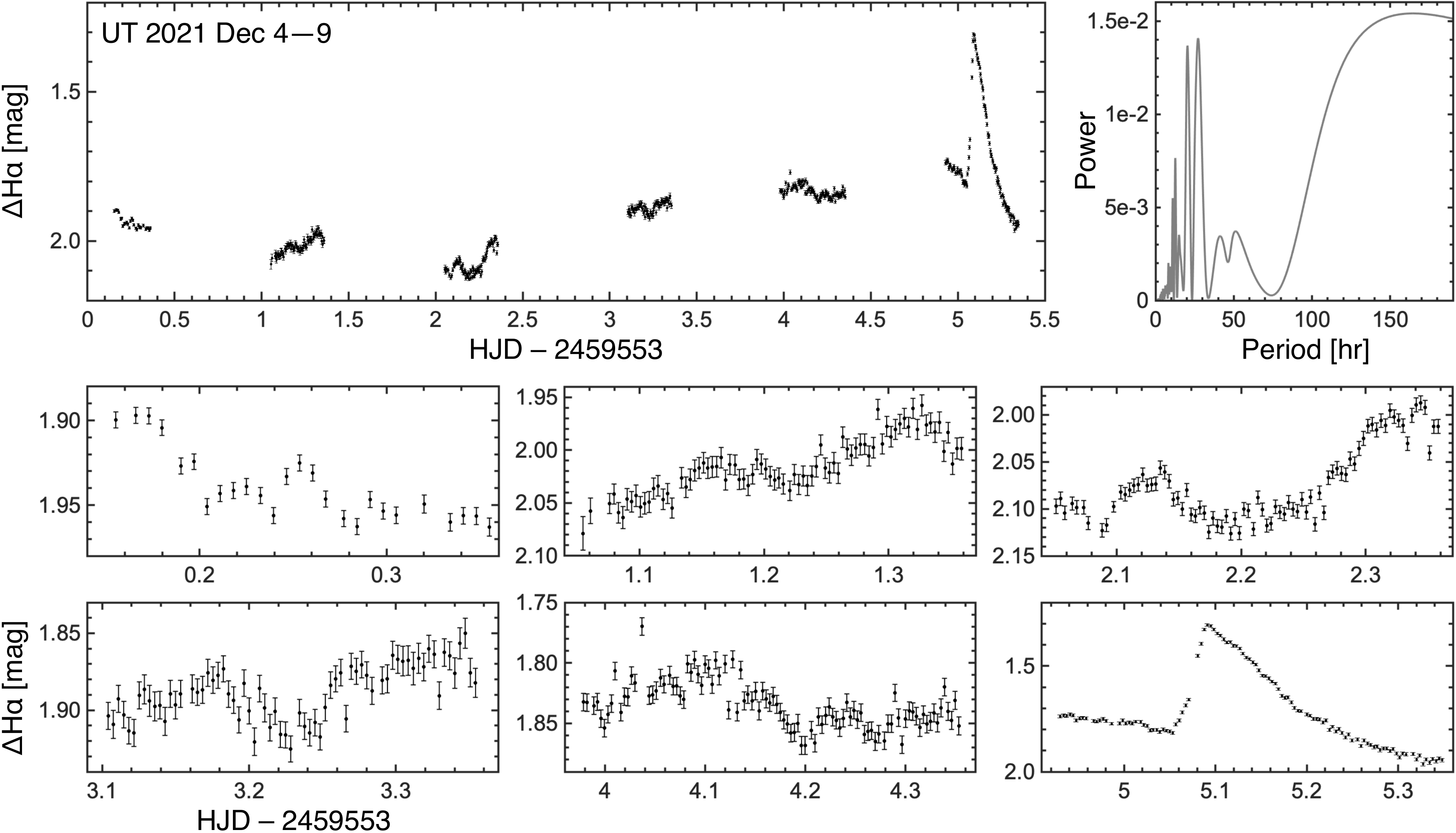}
\caption{Relative photometry between FU Tau A and 2MASS J04241447$+$2506106 between UT 2021 December 4--9. Top: H$\alpha$ light curve of FU Tau A and the corresponding Lomb--Scargle periodogram which peaks at 164.4 hr. Bottom: zoomed-in light curve of each night.} 
\label{fig:LC_A}
\end{figure*}

\begin{figure*}[h]
\centering
\figurenum{4}
\includegraphics[width=\linewidth]{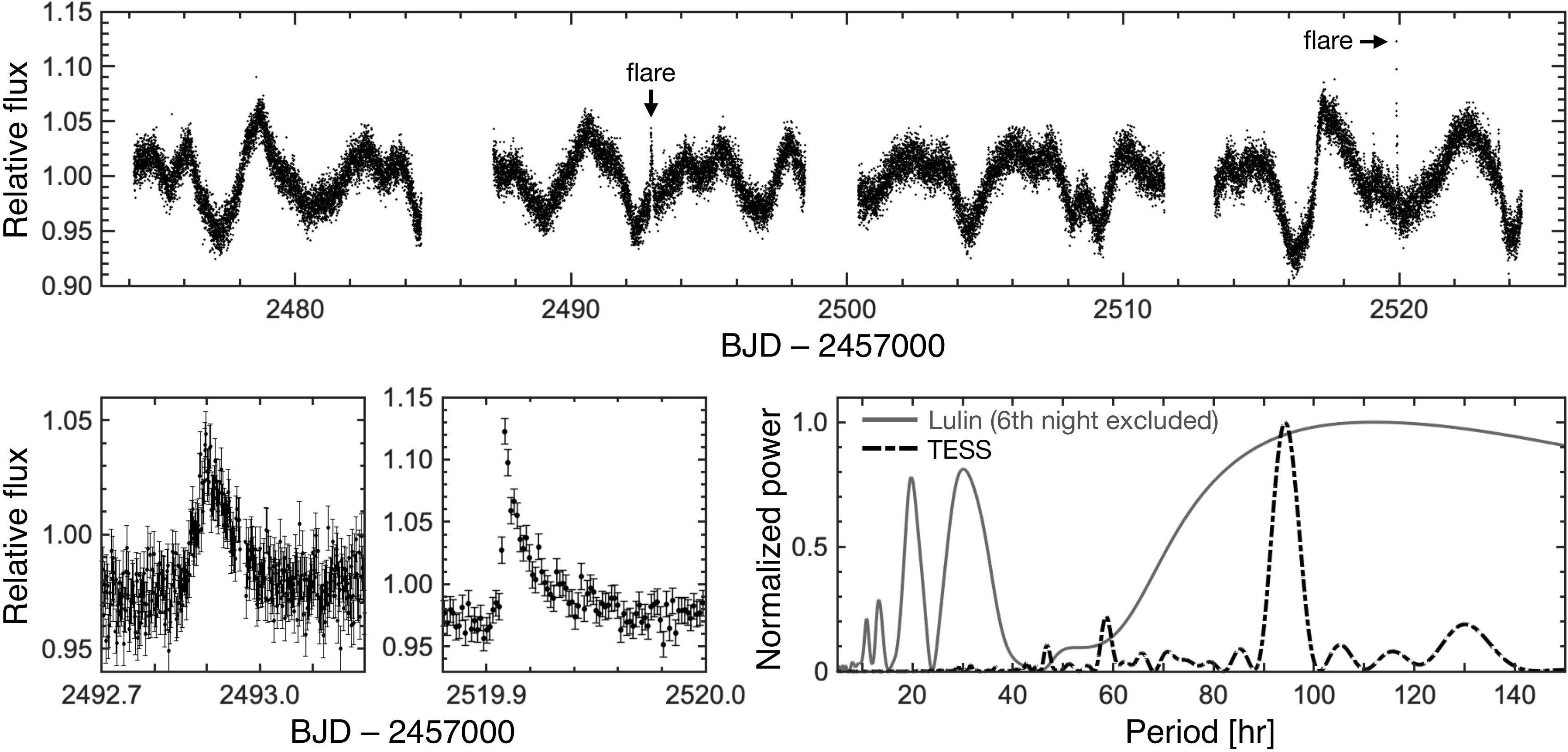}
\caption{Top: TESS light curve of FU Tau A (Sectors 43 and 44), which has been presented in \cite{Bowler23} and is reproduced here. Lower left: the two flares identified in the TESS data. Lower right: periodograms for the TESS and Lulin observations. When excluding the data on December 9, the Lulin curve peaks at 112.3 hr, much closer to the rotation period of 94.3 hr derived from the TESS data (see 164.4 hr in Figure \ref{fig:LC_A}), though also close to the duration of our monitoring.}
\label{fig:TESS}
\end{figure*}

\section{Results}
\label{results}
\subsection{Variable H$\alpha$ Emission from FU Tau A}
Figure \ref{fig:LC_A} displays the H$\alpha$ light curve of FU Tau A. On December 9, we detected a flare that rapidly brightened by about 0.5 mag in 50 minutes and gradually faded in the next 5 hr. We also search for flares in the TESS data by first dividing the original light curve into 1 day segments, from which we subtract a best-fit third-order polynomial and calculate the standard deviation of the residual curve. Flare candidates are selected as having at least three consecutive residual points greater than 3$\sigma$. We visually examine these candidates and identify two flares showing a characteristic rapid rise followed by an exponential decay (Figure \ref{fig:TESS}). Both flares, however, have weaker amplitudes of 7\% and 16\% and shorter durations of 5 hr and 1.7 hr, respectively. This is consistent with the standard scenario where white-light flares are shorter than chromospheric H$\alpha$ flares, reflecting different timescales between nonthermal and thermal heating (e.g., \citealt{Neupert68}). The TESS light curve clearly shows a periodic variation from which \cite{Bowler23} derived a rotation period of 94.3 hr, consistent with the ground-based measurements in \cite{Scholz12}. 

To further classify the light curve, we follow the approach in \cite{Cody14} and \cite{Cody18} to calculate the $M$ and $Q$ metrics. In brief, $M$ describes the asymmetry with respect to the mean and $Q$ reflects the periodicity. We find $M = 0.08$ and $Q = 0.23$, suggesting that FU Tau A is quasiperiodic symmetric, similar to 26\% of the Taurus young stars \citep{Cody22}. In addition, \cite{Cody18} showed that quasiperiodic sources tend to have inclined disks with $i>50$\degr. Although FU Tau A's disk was not spatially resolved with ALMA \citep{Wu20}, the tentative high disk inclination is in line with a high stellar inclination of $i_\ast=75^{+14}_{-5}\degr$ discovered by \cite{Bowler23}. 

The varying H$\alpha$ light curve suggests that accretion onto FU Tau A is constantly changing on minute-to-hour timescales. There appears to be a multiday sinusoidal variation with an amplitude of $\sim$0.15 mag that hints a rotational modulation. In the lower-right panel of Figure \ref{fig:TESS}, we compare the Lomb--Scargle periodograms \citep{Lomb76,Scargle82} derived from the TESS and the Lulin data. For the latter, we only include the first five nights to ensure that the derived period is not affected by the prominent flare. The resulting Lulin periodogram has a broad peak that overlaps the rotation period of 94.3 hr, but the peak is also close to the duration of our monitoring. Future H$\alpha$ monitoring that covers multiple rotation periods can examine if accretion onto FU Tau A is indeed rotationally modulated.

\subsection{Variable H$\alpha$ Emission from FU Tau B}
\label{phot_B}
Figure \ref{fig:LC_B} shows the H$\alpha$ light curve of FU Tau B. Albeit with large uncertainties of $\sim$0.1--0.2 mag, our observations suggest that the H$\alpha$ emission of FU Tau B may be variable on timescales of a few hours. There might be some sinusoidal variations in a number of nights, but it is not clear if there is any underlying periodicity that can be ascribed to the rotation period---the periodogram has similar power at around 9, 14, and 35 hr. Whether accretion onto FU Tau B is indeed rotationally modulated awaits more sensitive monitoring. We also note that no powerful accretion bursts or chromospheric flares are detected despite the fact that our monitoring potentially covers a timespan of multiple rotation periods.

\begin{figure}[t]
\centering
\figurenum{6}
\includegraphics[width=\linewidth]{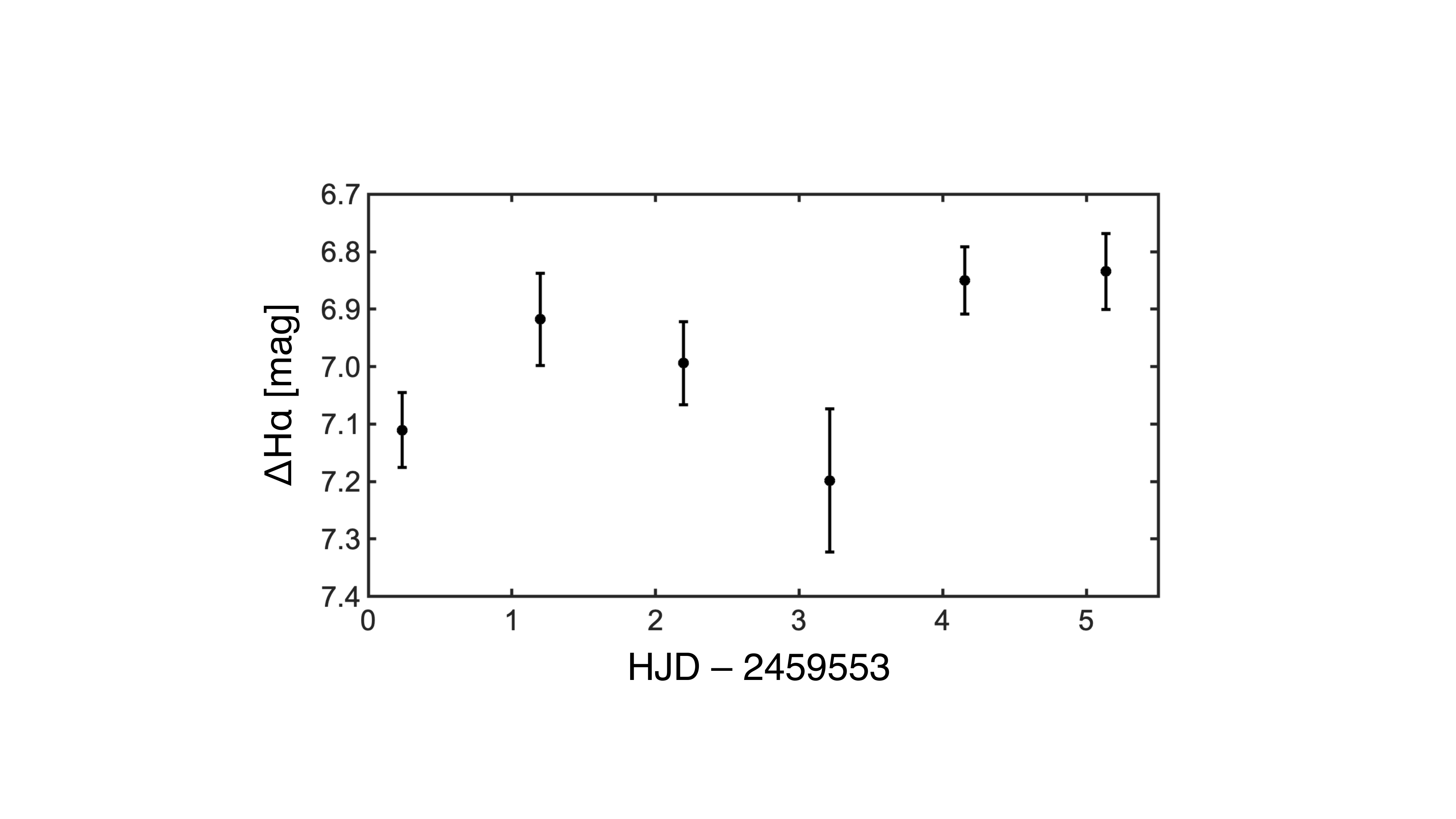}
\caption{Daily H$\alpha$ variation of FU Tau B. The data points are listed in Table \ref{tb4}.}
\label{fig:LC_B_day}
\end{figure}

In Figure \ref{fig:LC_B_day}, we examine H$\alpha$ variability on a daily timescale by merging all frames each night. Table \ref{tb4} lists $\Delta$H$\alpha$ and the uncertainties. FU Tau B appeared to be $\sim$0.3 mag brighter on December 8 and 9 than on December 4 and 7. Our observations hint a possible daily variation with a standard deviation of 0.15 mag. Overall, accretion onto FU Tau B seems to be largely stable with low-level hourly and daily variations, rather than drastically changing on short timescales.

\subsection{Converting H$\alpha$ Flux Density to an Accretion Rate}
\subsubsection{FU Tau B}
\label{sub_B}
While absolute calibration may be questionable due to the lack of optical standard stars, as 2MASS J04241447$+$2506106 appears stable to a $\lesssim$1\% level at H$\alpha$, here we assume that all of its detected H$\alpha$ comes from the photospheric continuum. We derive a spectral type of K7 by comparing its colors to \cite{Pecaut13}, an effective temperature of 4100 K by fitting the synthetic spectra in \cite{Coelho14} to the 2MASS and Gaia photometry, and a flux density of $2.36\times10^{-14}$ erg s$^{-1}$ cm$^{-2}$ \AA$^{-1}$ at 6563\,\AA \,assuming negligible extinction. 

\begin{figure*}[t]
\centering
\figurenum{5}
\includegraphics[width=\linewidth]{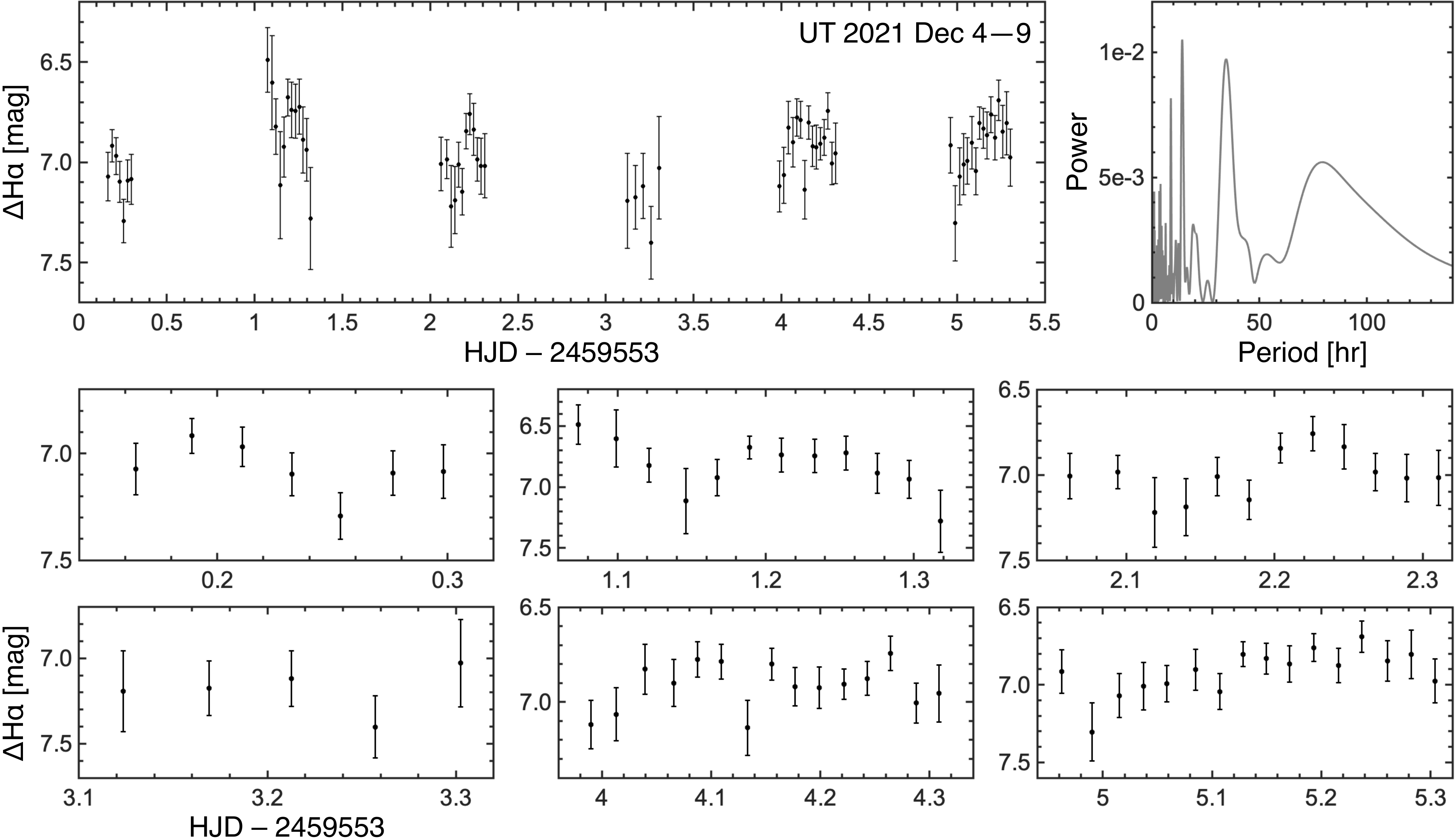}
\caption{Relative photometry between FU Tau B and 2MASS J04241447$+$2506106 between UT 2021 December 4--9. To increase the S/Ns, we merged the data into 30 minutes bins except for the fourth night, in which we merged into 1 hr bins. For reference, the variability for the binned reference star is about 0.3\%.}
\label{fig:LC_B} \vspace{5pt}
\end{figure*}

Since FU Tau B is on average 7 mag fainter (Figure \ref{fig:LC_B}), we estimate a dereddened H$\alpha$ flux density of $(5.4\pm1.1)\times10^{-17}$ erg s$^{-1}$ cm$^{-2}$ \AA$^{-1}$ assuming that FU Tau B shares the same extinction of $A_V = 0.5\pm0.5$ as FU Tau A \citep{Stelzer13} and propagating all of the uncertainties in a Monte Carlo fashion. This is $\sim$20 times higher than the $r$-band photometry in \cite{Luhman09} and \cite{Quanz10}, so we remove 5\% of the value above to account for the contribution from the adjacent continuum and yield $(5.1\pm1.1)\times10^{-17}$ erg s$^{-1}$ cm$^{-2}$ \AA$^{-1}$. This flux density corresponds to an H$\alpha$ line luminosity of $L_{\rm H\alpha} = (9.4\pm2.0)\times10^{-7}\,L_\odot$ by multiplying it by the effective filter width of 36.7\,\AA\,and $4\pi d^2$, where $d$ is the Gaia DR3 distance to FU Tau (see Section \ref{sect1}). Compared with other substellar companions near the deuterium-burning limit, FU Tau B has a $L_{\rm H\alpha}$ similar to that of SR 12 c ($6.5\times10^{-7}\,L_\odot$; \citealt{SantamariaMiranda19}) and DH Tau B ($6.5\times10^{-7}\,L_\odot$; \citealt{Zhou14}), but about 10--20 times weaker than that of GSC 06214--00210 B ($9.3\times10^{-6}\,L_\odot$; \citealt{Zhou14}) and GQ Lup B ($1.4\times10^{-5}\,L_\odot$; \citealt{Stolker21}). We also note that $L_{\rm H\alpha}$ of FU Tau B is similar to that of the embedded giant plant PDS 70 b ($6.5\times10^{-7}\,L_\odot$; \citealt{Zhou21}). Despite small number statistics, there is no clear correlation between $L_{\rm H\alpha}$ and the masses and ages of these accreting companions.

As we have no information of the shock-induced UV/optical continuum, we cannot directly measure the accretion luminosity $L_{\rm acc}$. As a result, we apply the $L_{\rm acc}$--$L_{\rm H\alpha}$ relationship in \cite{Aoyama21} to estimate $\log\,(L_{\rm acc}/L_\odot) =  -4.1 \pm 0.3$. We then follow \cite{Wu20} and derive a bolometric luminosity of $\log\,(L_{\rm bol}/L_\odot)=-2.51\pm0.05$ using the Gaia DR3 distance (see Section \ref{sect1}) and therefore a mass of $19\pm4$ \Mj \, by comparing $L_{\rm bol}$ with the model grids in \cite{Baraffe15}. Together with an effective temperature of 2400 K \citep{Luhman09} and a nominal uncertainty of 100 K, we estimate a radius of $3.2\pm0.3 \,R_{\rm Jup}$. Substituting these numbers and the accretion luminosity into the magnetospheric prescription, $\dot{M}=1.25\frac{R_\ast  L_{\rm acc}}{GM_\ast}$, in \cite{Gullbring98}, we calculate an accretion rate of $(5.9\pm4.3)\times10^{-8}$ \Mj \,yr$^{-1}$ for FU Tau B. If considering the scatter in $\Delta$H$\alpha$ (6.5 to 7.4 mag), the median value of the accretion rate is within the range of (4--9)$\times10^{-8}$ \Mj \,yr$^{-1}$. A value of $10^{-7}$--$10^{-10}$ \Mj \,yr$^{-1}$ has been commonly reported for young planetary-mass objects (e.g., \citealt{Zhou14,Haffert19,SantamariaMiranda19,Eriksson20}). We caution that our accretion rate estimate is likely a lower limit because we have no information about the intrinsic H$\alpha$ flux density from 2MASS J04241447$+$2506106 or there could be substantial reddening from the disk around FU Tau B (but see discussions of negligible dust extinction at H$\alpha$ in \citealt{Marleau22}). 
\vspace{-10pt}

\begin{deluxetable}{ccc}
\tablewidth{\columnwidth}
\tablecaption{Daily H$\alpha$ Variation of FU Tau B\label{tb4}}
\tablehead{
\colhead{\hspace{.82cm} Date}\hspace{.82cm}  &
\colhead{\hspace{.82cm} $\Delta{\rm H}_\alpha$}\hspace{.82cm}  &
\colhead{\hspace{.82cm} $\sigma_{\Delta{\rm H}\alpha}$}\hspace{.82cm}  
}
\startdata 
2459553.23612 & 7.11 & 0.07 \\ 
2459554.19870 & 6.92 & 0.08 \\ 
2459555.19387 & 6.99 & 0.07 \\ 
2459556.21297 & 7.20 & 0.13 \\ 
2459557.15304 & 6.85 & 0.06 \\ 
2459558.13737 & 6.83 & 0.07
\enddata
\tablenotetext{}{{\bf Note.} (1) Mean heliocentric Julian date. (2) Magnitude difference at H$\alpha$. (3) Uncertainty of the magnitude difference at H$\alpha$.} \vspace{-20pt}
\end{deluxetable}

To understand whether the observed H$\alpha$ from FU Tau B can be mostly chromospheric, we follow the empirical relationship for young brown dwarfs in \cite{Manara13,Manara17} to estimate the chromospheric contribution to the accretion luminosity as log $(L_{\rm acc,noise}/L_{\rm bol}) = (6.2\pm0.5)\times\log \,(T_{\rm eff})-(24.5\pm1.9)$. Substituting log $(L_{\rm bol}/L_\sun)$ and $T_{\rm eff}$ for FU Tau B, we find $L_{\rm acc,noise} \sim 10^{-6}\,L_\sun \ll L_{\rm acc} \sim 10^{-4.1}\,L_\sun$. Therefore, the observed H$\alpha$ emission from FU Tau B likely originates from mass accretion. We also note that the observed H$\alpha$ equivalent width of 70\,\AA\,\citep{Luhman09} is higher than that of accreting brown dwarfs in \cite{Natta04} and also the M8.5 SSSPM J1102--3431 in \cite{Herczeg09}.

\subsubsection{FU Tau A} \vspace{-2pt}
\label{sub_A}
Figure \ref{fig:LC_A} shows that FU Tau A is on average 1.94 mag fainter than the reference star when excluding data of the sixth night. We hence derive an H$\alpha$ flux density of $(5.7\pm1.4)\times10^{-15}$ erg s$^{-1}$ cm$^{-2}$ \AA$^{-1}$, about an order of magnitude higher than the $r$-band photometry in \cite{Luhman09}, so we remove 10\%~of the value above to account for the contribution from the adjacent continuum and yield $(5.1\pm1.2)\times10^{-15}$ erg s$^{-1}$ cm$^{-2}$ \AA$^{-1}$. The H$\alpha$ line luminosity is then $L_{\rm H\alpha} = (9.4\pm2.2)\times10^{-5}\,L_\odot$. We then apply the $L_{\rm acc}$--$L_{\rm H\alpha}$ relationship in \cite{Rigliaco12} to estimate $\log\,(L_{\rm acc}/L_\odot) = -3.0 \pm 0.3$.

To estimate the mass of FU Tau A, we adopt the revised bolometric luminosity of log $(L_{\rm bol}/L_\sun) = -0.96\pm0.05$ in \cite{Bowler23} and compare with the model grids in \cite{Baraffe15}. Our analysis favors a low system age of $\sim$1 Myr and a somewhat higher mass of $0.14\pm0.03$ $M_\sun$ than 0.05 $M_\sun$ in \cite{Luhman09} and 0.08 $M_\sun$ in \cite{Stelzer13}. Together with a stellar radius of $1.4\pm0.1\,R_\sun$ \citep{Bowler23}, we estimate a mean accretion rate of $\dot{M}=(4.0\pm2.9)\times 10^{-10}\,M_\sun$\,yr$^{-1}$. This value is similar to the previous measurements of $3.5\times 10^{-10}$\,$M_\sun$\,yr$^{-1}$ in \cite{Stelzer10} and $1.3\times 10^{-10}$\,$M_\sun$\,yr$^{-1}$ in \cite{Stelzer13}, and is about 7 times higher than that of FU Tau B ($\sim$$6\times10^{-11}$\,$M_\sun$\,yr$^{-1}$; Section \ref{sub_B}).

\section{Discussion}
While magnetospheric accretion has been highly successful and observationally confirmed in the stellar regime (e.g., \citealt{Gravity20}), it remains unclear whether young planetary-mass objects (giant planets, free-floating planets, brown-dwarf companions, etc.) can have a strong magnetic field to truncate their disks and trigger magnetospheric accretion. \cite{Owen16} argued that if the field is weaker than 65 G, planetary accretion may follow the alternative boundary-layer accretion scheme (e.g., \citealt{Szulagyi20,Takasao21}), where the circumsubstellar disk would directly touch the planetary surface and form a hot \textit{ring} with a surface filling factor likely 10 times higher than that in magnetospheric accretion. \cite{Takasao21} showed that in this case the line intensity could fluctuate due to the unstable accretion streams, but the line profile would be insensitive to planetary rotation. Therefore, detecting a periodically changing line profile could be a telltale sign of the magnetospheric models. For young stars and brown dwarfs, rotational modulation in the line profiles has been detected (e.g., \citealt{SJ06,Stelzer07,Donati08,Costigan14,Alencar18,Pouilly20}). Recently, \cite{Ringqvist23} resolved the hydrogen lines for the $\sim$13\,\Mj\,circumbinary companion Delorme 1 (AB)b, and they derived a filling factor of 1\% for the line-emitting area, smaller than $\mathcal{O}(0.1)$ in boundary-layer accretion (see \citealt{Takasao21}). Along with the asymmetric line profiles nicely fit with multiple velocity shifts and widths, these features are consistent with the magnetospheric accretion models. A $<$1\% filling factor has also been inferred for the two 10--30 $M_{\rm Jup}$ wide companions GQ Lup B and GSC 06214--00210 B \citep{Demars23}.

Accretion onto brown dwarfs can be variable on different timescales (e.g., \citealt{Natta04,SJ06,Stelzer07}), sometimes with amplitudes varying by orders of magnitude (e.g., \citealt{Nguyen-Thanh20}). Existing H$\alpha$ observations for companions near the planet-brown-dwarf boundary, though mostly sparsely sampled and with large uncertainties, seem to find low-level variations. For Delorme 1 (AB)b, \cite{Eriksson20} found that the H$\alpha$ line luminosities remained stable within 1--2$\sigma$ ($\sigma\sim$ 30--40\%) over 2.5 hr. For PDS 70 b, \cite{Zhou21} showed that their six-epoch HST H$\alpha$ photometry ($\sigma \sim$ 20\%) was consistent within $\lesssim$3$\sigma$ with previous adaptive optics measurements \citep{Wagner18,Haffert19,Hashimoto20}. While our LOT monitoring of FU Tau B lacks a sufficient sensitivity to reveal any rotational modulation, we see no large-amplitude H$\alpha$ variations for six nights of consecutive observations, implying that accretion onto FU Tau B may be relatively stable on timescales of hours to days.

Nonetheless, accretion variability has been found to increase on longer timescales for young stars (e.g., \citealt{Nguyen09,Mendigutia11,Pogodin12,Costigan14,Zsidi22}), brown dwarfs (e.g., \citealt{Stelzer07}), and even for substellar companions \citep{Demars23}. \cite{Stelzer07} found that accretion onto the 20--30~\Mj~brown dwarf 2MASS J12073346--3932539 mildly varied between $10^{-7.1}$ and $10^{-6.8}$\,\Mj\,yr$^{-1}$ (a factor of two) for two consecutive nights, but changed by at least a factor of 10 on monthly to yearly timescales. \cite{Wolff17} also found that the equivalent width of the 1.282~\micron~Pa$\beta$ emission from the 10--15\,\Mj\,DH Tau B could vary by a factor of a few in weeks. Recently, \cite{Demars23} show that the variability amplitude of the Pa$\beta$ line stays low ($<$50\%) on hourly timescales but becomes significant ($\gtrsim$100\% and up to $\sim$1000\%) on monthly to yearly timescales for GQ Lup B and GSC 06214--00210 B. To examine if FU Tau B exhibits the same feature, we follow \cite{Demars23} to calculate the flux variation as (flux$_{\rm max}-$flux$_{\rm min}$)/flux$_{\rm min}$ for each individual pair of measurements in Table \ref{tb3}, where flux$_{\rm max}$ and flux$_{\rm min}$ are the higher and the lower values of each pair, respectively. Due to our large photometric uncertainties, however, we find a similar flux variations of 10\%--20\% (median) from hourly to daily timescales. Multiepoch H$\alpha$ monitoring is essential to reveal if the variability amplitude would be larger on longer timescales for FU Tau B and more substellar companions. On the other hand, \cite{Claes22} find that accretion variability estimated with UV continuum excess can be $\sim$1 dex higher than that with emission line luminosity. It would be illuminating to carry out extensive UV monitoring to ascertain the true variability. Finally, long-term monitoring over several Keplerian timescales of the inner circumsubstellar disk would constrain the frequency of bursts and dips, offering insights into the structures of the disk as well as the accretion funnels.

\section*{Acknowledgements}
\noindent We thank the anonymous reviewer for the very constructive comments. Y.-L.W. acknowledges the support from the National Science and Technology Council (grant 111-2636-M-003-001) and the Ministry of Education. This publication has made use of data collected at Lulin Observatory, partly supported by NSTC grant 109-2112-M-008-001. B.P.B. acknowledges support from the National Science Foundation grant AST-1909209, NASA Exoplanet Research Program grant 20-XRP20$\_$2-0119, and the Alfred P. Sloan Foundation. This research has made use of the NASA/IPAC Infrared Science Archive, which is funded by the National Aeronautics and Space Administration and operated by the California Institute of Technology. The corresponding DOI for the Zwicky Transient Facility Image Service is:\dataset[10.26131/irsa539]{https://doi.org/10.26131/irsa539}. Funding for the TESS mission is provided by NASA's Science Mission directorate. This paper includes data collected by the TESS mission, which are publicly available from the Mikulski Archive for Space Telescopes (MAST)\dataset[10.17909/aq6f-df31]{https://doi.org/10.17909/aq6f-df31}. Finally, this research has made use of MATLAB R2020b.

\facilities{LO:1m, IRSA.}
\software{MATLAB.}

\end{document}